# Strong thermal transport along polycrystalline transition metal dichalcogenides revealed by multiscale modelling for MoS$_2$

Bohayra Mortazavi[*,1], Romain Quey[2], Alireza Ostadhossein[3], Aurelien Villani[2], Nicolas Moulin[2], Adri C. T. van Duin[4], Timon Rabczuk[#,5]

[1]Institute of Structural Mechanics, Bauhaus-Universität Weimar,

Marienstr. 15, D-99423 Weimar, Germany.

[2]École des Mines de Saint-Étienne, CNRS UMR 5307, 158 cours Fauriel,

F-42023 Saint-Étienne, Cedex 2, France.

[3]Department of Materials Science and Engineering, University of Pennsylvania,

Philadelphia, Pennsylvania 19104, USA.

[4]Department of Mechanical and Nuclear Engineering, Pennsylvania State University,

University Park, USA.

[5]College of Civil Engineering, Department of Geotechnical Engineering,

Tongji University, Shanghai, China.

## Abstract

Transition metal dichalcogenides (TMDs) represent a large family of high-quality 2D materials with attractive electronic, thermal, chemical, and mechanical properties. Chemical vapour deposition (CVD) technique is currently the most reliable route to synthesis few-atomic layer thick and large-scale TMDs films. However, the effects of grain boundaries formed during the CVD method on the properties of TMDs nanomembranes have remained less explored. In this study, we therefore aim to investigate the thermal conduction along polycrystalline molybdenum disulfide (MoS$_2$) as the representative member of TMDs nanomembranes family. This goal was achieved by developing a combined atomistic-continuum multiscale method. In the proposed approach, reactive molecular dynamics simulations were carried out to assess thermal contact conductance of diverse grain boundaries with various defects configurations. The effective thermal conductivity along the CVD grown polycrystalline and single-layer MoS$_2$ was finally acquired by conducting finite element modelling. Insight provided by this investigation can be useful to evaluate the effective thermal transport along a wide variety of 2D materials and structures.

*Corresponding author (Bohayra Mortazavi):  bohayra.mortazavi@gmail.com

Tel: +49 157 8037 8770, Fax: +49 364 358 4511, [#]Timon.rabczuk@uni-weimar.de

# 1. Introduction

First mechanical exfoliation of graphene [1-4] from graphite, raised an increasing attention toward two-dimensional (2D) materials [1-5] as a new class of materials with outstanding properties, well-suited for diverse applications ranging from nanoelectronics to aerospace structures. Since then, tremendous experimental efforts have been devoted to synthesize extensive types of 2D materials. As a results of experimental achievements during the last decade, currently a wide variety of few-atomic layer thick, high-quality and large-scale 2D materials are available such as hexagonal boron-nitride (h-BN) [6,7], graphitic carbon nitride [8,9], silicene [10,11], phosphorene [12,13], borophene [14], germanene [15], stanene [16] and transition metal dichalcogenides (TMDs) like $MoS_2$, $WS_2$, $MoSe_2$ and $WSe_2$ [17,18]. Although the graphene presents the highest known mechanical [19] and thermal conduction [20] properties, its zero-band-gap semiconducting electronic character limit its suitability for many applications. In this regard, $MoS_2$, $WS_2$, $MoSe_2$ and $WSe_2$, TMDs films [21-27] have attracted remarkable attentions because of presenting direct band-gap in their single-layer forms, which propose them as promising building-blocks, such as transistors, photodetectors and electroluminescent in nanodevices [21,28,29].

Reliable and large-scale synthesis of TMDs nanomembranes with homogenous properties plays however an essential role for their practical application in nanodevices. Now-a-days, TMDs films are mainly synthesized using the chemical vapour deposition (CVD) or chemical exfoliation techniques [21]. For the synthesis of large-area graphene [30-32] or h-BN [33-36] nanosheets, CVD method has been employed as the most promising technique. In the similar way, CVD method has been also effectively used to fabricate large-scale and high-quality atomically thin TMDs nanomembranes [37-45]. An interesting fact about the TMDs films is that CVD method can be employed to fabricate both lateral and in-plane heterostructures [46-49] made from different TMDs structures in which the electronic and optical properties are further tuneable. The CVD method is therefore currently among the most promising routes toward the synthesis of large-area 2D materials, so that the properties of CVD grown nanomembranes should be studied in detail. It is worthy to mention that during the CVD process, the crystal growth of individual grains often leads to the fabrication of polycrystalline structures. In the polycrystalline nanomembranes, at the places where the grains with different growth orientations meet, grain boundaries form. These grain boundaries extend throughout the structure and due to their different atomic lattices, they are distinguishable from the atoms inside the pristine grains, those only with hexagonal atomic structure. Since the hexagonal atomic lattice is the most stable structure for $MoS_2$, $WS_2$, $MoSe_2$, $WSe_2$, $MoTe_2$ and $WTe_2$, the non-hexagonal atomic lattices existing along the grain boundaries can be considered as topological defects. These defects can act as dislocation cores that originate stress concentration, scatter the phonons and alter the electronic states [50] along the grain boundaries and such that they may accordingly decline mechanical [51,52] and thermal conduction properties [53-55] or substantially affect the electronic properties [56-58]. In this case by decreasing the grain size, more grain boundaries form in the CVD grown films and the density of defects consequently increases.

Among the TMDs nanomembranes fabricated so far, molybdenum disulfide ($MoS_2$) [18,21] has attracted the most attention. $MoS_2$ presents semiconducting electronic character with a direct band-gap of 1.8 eV which has been experimentally produced in free-standing and single-layer form. Among the various properties of 2D materials, their thermal conductivity plays an import role in the design of nanodevices. In the applications related to the nanoelectronics and energy storage, to minimize the overheating issues, the thermal conductivity of employed 2D material should be high enough to efficiently and promptly transmit the heat to the environment in order to avoid failure of the system due to the excessive heats. For example, $MoS_2$ thin-film transistors [59,60] have applications in high-temperature electronics [61] and in this case their thermal conductivity plays a crucial role not only in their performance but also for the safety and durability of the nanodevices. On the other hand, for particular applications like thermoelectricity, a 2D material with a lower thermal conductivity is more advantageous to improve efficiency in the energy conversion. By taking into consideration that in many application, TMDs nanomembranes are in polycrystalline form, a fundamental understanding of the thermal transport along the polycrystalline TMDs nanomembranes is of a crucial importance. In this investigation, our objective is to explore the heat conduction along the polycrystalline $MoS_2$ as the representative member of TMDs nanosheets. To this aim, we developed a combined atomistic-continuum multiscale modelling. We discuss that the proposed hierarchical multiscale approach can be considered to efficiently analyze the effective thermal conductivity of various 2D nanomembranes and heterostructures with complex atomic structures and grain boundaries.

## 2. Multiscale modelling

Based on the high resolution electron microscopy observations, it was found that MoS$_2$ grain boundaries, contain a wide variety of dislocation cores, consisting of pentagon-heptagon (5-7), tetragon- tetragon (4-4), tetragon-hexagon (4-6), tetragon-octagon (4-8), and hexagon-octagon (6-8) rings [43–45]. These experimental observations supported by the first-principles calculations [43–45], and clearly reveal more complex configurations for grain boundaries in MoS$_2$ than in graphene or h-BN [32,36], for which the grain boundaries mainly include pentagon-heptagon dislocation cores.. Interestingly, it was experimentally observed and theoretically confirmed that only for the pentagon-heptagon defect pair along the MoS$_2$ grain boundaries, homo-nuclear S-S bonds can form [44,45]. Fig. 1, depicts the atomic structure of 20 different grain boundaries that were constructed in this study according to the experimental observations for MoS$_2$ [43–45]. If the two MoS$_2$ grains that meet each other are both oriented along the armchair direction, the resulting grain boundary presents 4-4 or 4-8 rings. On another side, if both grains are oriented along the zigzag direction, the formed grain boundary includes a 4-8 dislocation core. Nonetheless, for 5-7, 4-6 and 6-8 rings, there exist many possibilities for the MoS$_2$ grain boundary configurations. Interestingly, for two MoS$_2$ grains with a particular misorientation angle, from the geometrical point of view, for the constructed grain boundary, there exist the possibility of formation of all 5-7, 4-6 or 6-8 dislocation cores simultaneously depending on the distances between the atoms in the two sides of the grain boundary. We note that the phonon scattering rates along the grain boundaries correlate with their defect concentrations in such a way higher thermal resistance is expected as the defect concentration increases. Among numerous possibilities for the construction of grain boundaries, we therefore selected 6 cases in which the defect concentration gradually increases. As shown in Fig. 1 for 5-7 grain boundaries, the defect concentration along the grain boundaries gradually changes and for the most and least defective cases there exist no hexagonal ring and 5 hexagonal rings, respectively, separating the two 5-7 dislocation cores. In a consistent manner, 6-8 and 4-6 grain boundaries were also constructed in a similar way as that we used to construct those with 5-7 rings. In these cases, however, the lateral mismatch between the atoms from the two sides of the grain boundary are different. In our modelling, we considered both the symmetrical and non-symmetrical grain boundaries, as shown in Fig. 1 for 5-7 dislocation cores. We also note that all created structures in this work are periodic along the grain boundary direction.

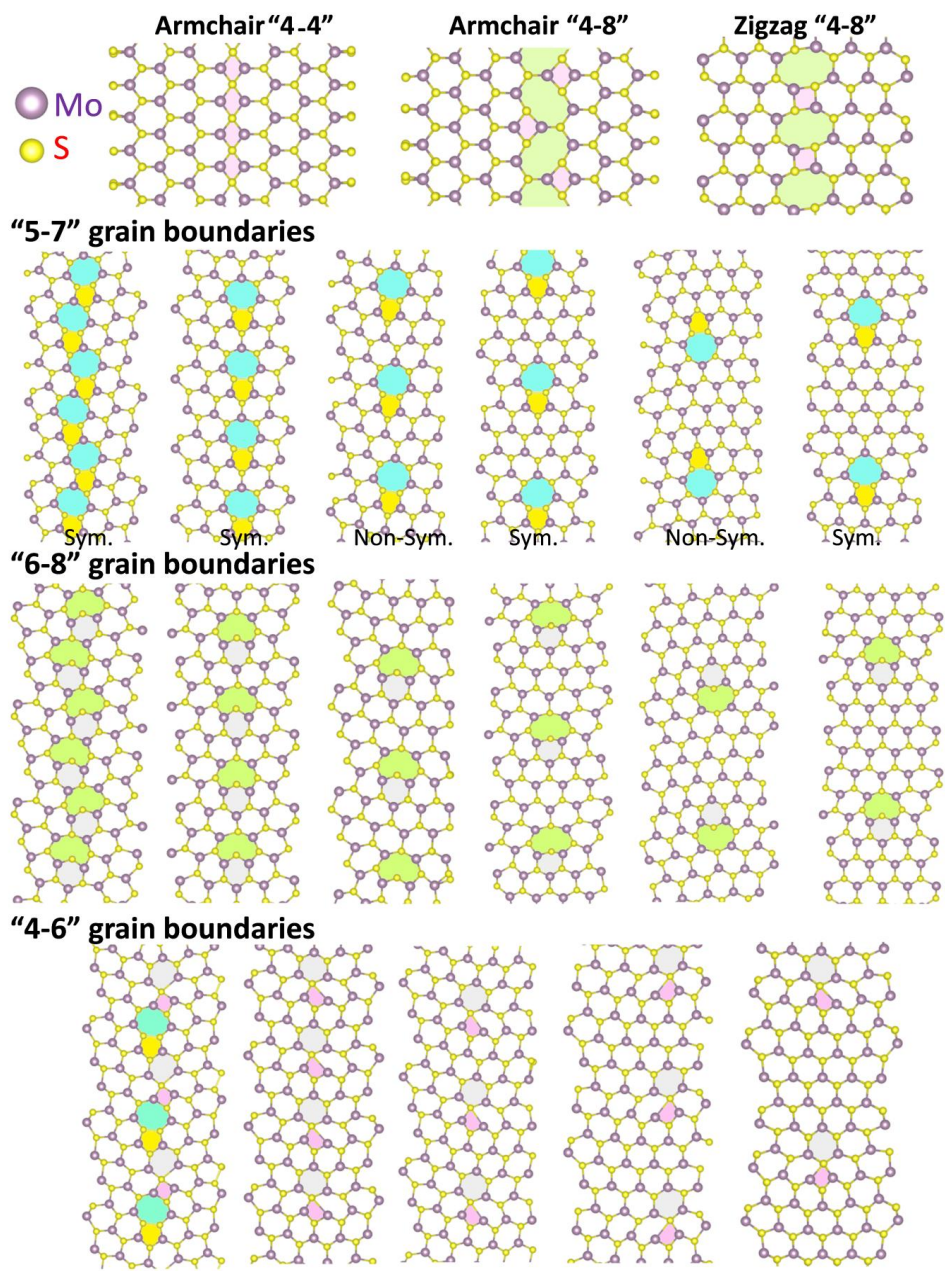

Fig. 1, Atomic configuration of 20 different MoS$_2$ grain boundaries that were constructed on the basis of experimental observations [43–45]. Armchair grain boundaries present 4-4 or 4-8 dislocation cores and zigzag grain boundary reveals 4-8 rings. For grain boundaries with 5-7, 4-6 and 6-8 dislocation cores, we selected few cases in which the defect concentration gradually increases or in another word the number of separating hexagonal rings decreases. In this case, we considered the both symmetric and non-symmetric grain boundaries as illustrated for the case of 5-7 defect pairs. All constructed structures are periodic along the grain boundary direction.

In this study, we explore the thermal conductivity along MoS$_2$ polycrystalline films through developing a combined atomistic-continuum (NEMD/FE) multiscale method. We first used reactive non-equilibrium molecular dynamics (NEMD) simulations not only to compute the thermal conductance of the 20 constructed MoS$_2$ grain boundaries but also to obtain the heat conductivity of pristine MoS$_2$. Based on the information provided by the classical NEMD simulations, we then constructed continuum models of polycrystalline MoS$_2$ using the finite element method to finally evaluate the heat transfer and effective thermal conductivity of samples at macroscopic scale. To further ensure the validity of our proposed technique, we also used similar multiscale approach to evaluate the thermal conductivity along polycrystalline graphene and h-BN and then we compared the results with effective thermal conductivity of fully atomistic models obtained using the equilibrium molecular dynamics (EMD) method.

Molecular dynamics calculations in this study were performed using LAMMPS (*Large-scale Atomic/Molecular Massively Parallel Simulator*) [62] package. The interaction of MoS$_2$ atoms were simulated using the ReaxFF bond order potential [63]. The energy of an atomic system ($E_{system}$) in ReaxFF is described by:

$$E_{\text{system}} = E_{\text{bond}} + E_{\text{val}} + E_{\text{tor}} + E_{\text{over}} + E_{\text{under}} + E_{\text{lp}} + E_{\text{vdW}} + E_{\text{coulombic}} \quad (1)$$

as the sum of the bond energy ($E_{bond}$), valence-angle ($E_{val}$), torsion-angle ($E_{tor}$), lone pair ($E_{lp}$), over-coordinate ($E_{over}$) and under-coordinate ($E_{under}$), energy penalties ($E_{lp}$), plus the nonbonded van der Waals ($E_{vdw}$) and Coulomb ($E_{coulombic}$) contributions. In our recent study [64], we studied the mechanical properties of all-MoS$_2$ heterostructures and confirmed the accuracy of the ReaxFF in simulating their mechanical responses. NEMD method was employed to study the thermal conduction properties. In this case, the time increment of simulations was set to 0.25 fs. The NEMD modelling conducted in this study is very similar to those we used in our recent works [65]. In the NEMD approach, we first equilibrated the structures at room temperature (300 K). Then, atoms at the two ends were fixed and the system was further equilibrated using the Nosé-Hoover thermostat method (NVT). We partitioned the simulation box into 22 slabs and applied a temperature difference of 20 K only between the first and last slabs [65]. In this step, the temperatures for these two slabs, so called hot (310 K) and cold (290 K) slabs were controlled at the adjusted values using the NVT method while the rest of the structure (i.e., the remaining 20 slabs) was simulated without any temperature control using the constant energy (NVE) method. As a result of applied loading condition, the steady-state heat transfer condition is achieved and results in a constant heat-flux, $J_x$, in the structure together with a temperature profile along the specimen length. We note that we assumed a thickness of 6.1 Å for single-layer $MoS_2$ [64]. For the thermal conductivity of $MoS_2$ single-layers without a grain boundary, a linear temperature relation ($dT/dx$=constant) was established along the sample. Based on the applied heat-flux and the established temperature gradient, the thermal conductivity, k, of pristine $MoS_2$ can then be calculated using the one-dimensional form of the Fourier law:

$$k = J_x \frac{dx}{dT} \qquad (2)$$

For the evaluation of $MoS_2$ grain boundaries thermal conductance, we calculated the established temperature difference, $\Delta T$, at the middle of sample, because of the grain boundary existence. The grain boundary conductance, $k_{GB}$, can be calculated using the following relation:

$$k_{GB} = \frac{J_x}{\Delta T} \qquad (3)$$

Finally, the effective thermal conductivity of polycrystalline $MoS_2$ as a function of grain size, was obtained using the finite element (FE) method. The microstructure was generated using the Neper software package [66], as a Laguerre tessellation [67]. A 2-D Laguerre tessellation of a domain of space, D, is constructed from a set of N seeds, $S_i$ (i ϵ 2 [1; N]), of specific positions, $x_i$, and non-negative weights, $\omega_i$. To each seed ($S_i$) is associated a cell, $C_i$, as follows:

$$C_i = \{P(x) \in D \mid d_E(P, S_i)^2 - w_i < d_E(P, S_j)^2 - w_j \;\; \forall i \neq j\} \qquad (4)$$

where $d_E$ is the Euclidean distance. The seed properties were set by an optimization approach in order for the resulting tessellation cells to match specified morphological properties. The tessellation properties are imposed using two metrics to characterize the cells. The distribution of cell size is controlled using the so-called relative "equivalent diameter", $d_r$, which corresponds to the ratio between the diameter of the circle of equivalent area of a cell and the average value over all cells. The distribution of the cell shapes is controlled using the so-called "circularity", c, which corresponds to the ratio between the perimeter of the circle of equivalent of the cell and the perimeter of the cell and takes the maximal value of 1 for a circular cell. In this work, the relative grain size distribution was represented by a lognormal law of standard deviation 0.45 (average 1), while the grain circularity distribution was represented by a normal distribution of average 0.9 and standard deviation 0.03. In our recent work [68] we showed that structures made on the basis of lognormal functions can well reconstruct the experimental samples of polycrystalline $MoS_2$ multi-layer nanomembranes. A microstructure containing 10,000 grains was generated in this way and then meshed into linear triangular elements, with an average of 60 elements per grain, to ensure the accuracy of finite element simulations. Grains were considered as independent (but still adjacent) bodies, with a discontinuous mesh at boundaries to model heat transfer resistance existing between connecting grains. In this case, NEMD results for grain boundaries thermal conductance were randomly assigned to every grain boundary to define the properties of contact elements. Finally, heat transfer was simulated by imposing a temperature difference of $\Delta T = 1$ K to two opposite sides of the domain, while the resulting temperature and thermal flux fields, were computed by the finite element method, using the Z-set software package [69-71]. The effective thermal conductivity of the polycrystalline films was finally acquired on the basis of established average heat-flux along the applied temperature difference, $h_f$, length of the constructed polycrystalline sheet, $L$, and applied temperature difference, $\Delta T$, using the following relation:

$$k_{eff} = h_f \frac{L}{\Delta T} \qquad (5)$$

## 3. Results and discussions

We first investigate the thermal conductivity of pristine and single crystalline $MoS_2$. In this case, the NEMD simulations were carried out for samples with different lengths to explore the length effect on the thermal conductivity. An increasing trend in the thermal conductivity value predicted by the NEMD was observed as the length of the sample increases. Nevertheless, the thermal conductivity of $MoS_2$ with infinite length can be calculated by the extrapolation of the NEMD results for the samples with finite lengths [72]. In Fig. 2a, the NEMD results for the inverse of thermal conductivity of pristine $MoS_2$ as a function of length inverse are illustrated. By extrapolation of the NEMD results for finite lengths, the thermal conductivity of pristine $MoS_2$ with an infinite length at room temperature was calculated to be 37±3 W/mK. Based on the experimental reports, the lattice thermal conductivity of few-layer thick $MoS_2$ were reported to be 34.5 ± 4 W/mK [73] and ~52 W/mK [74]. Nonetheless, the thermal conductivity of high-quality bulk $MoS_2$ films were experimentally measured to be in a range of 85–110 W/mK [75]. Interestingly, our NEMD modelling result based on the ReaxFF forcefield falls within the wide range of experimental reports of 34.5-110 W/mK [73-75] for the in-plane thermal conductivity of $MoS_2$ films. We however emphasize that our objective is to comparatively explore the effect of grain size on the thermal transport along polycrystalline MoS$_2$ films rather than to report exact thermal conductivity values. On the other side, concerning the thermal conductance along the MoS$_2$ grain boundaries, we found that because of the dominance of the phonon-defect scattering rate along the grain boundaries, the length effect in their thermal conductance is acceptably negligible (see Fig. 2b).

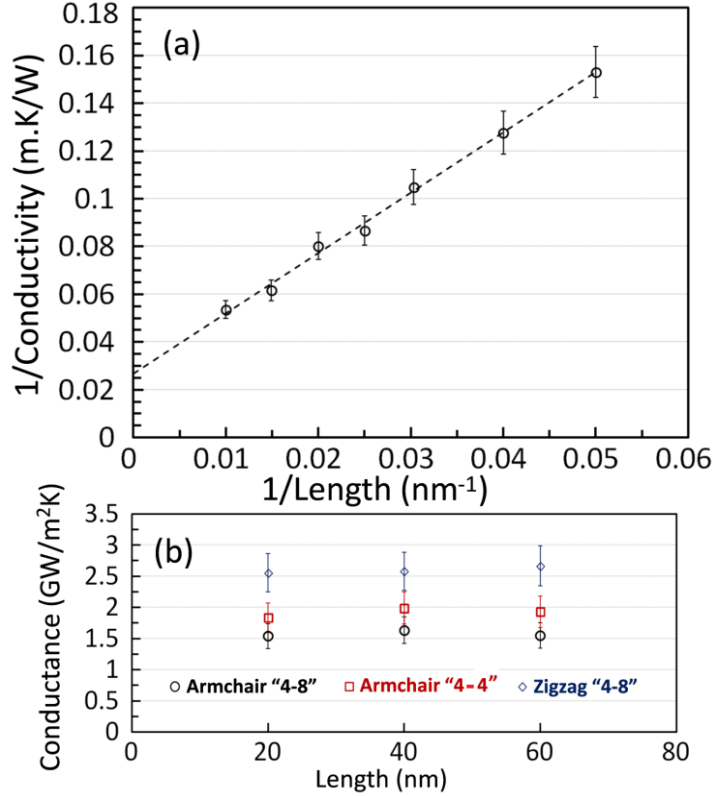

Fig. 2, (a) NEMD results for the calculated inverse of thermal conductivity of single-layer and defect-free MoS$_2$ as a function of length inverse. By fitting a linear curve (as shown here with the dashed line) and extrapolation to the zero point for the length inverse, the thermal conductivity of pristine MoS$_2$ with infinite length was acquired. (b) Computed thermal conductance of several grain boundaries for 3 different lengths revealing that the length effect on the thermal conductance of MoS$_2$ grain boundaries is negligible.

In Fig. 3, the NEMD predictions for the thermal contact conductance of various MoS$_2$ grain boundaries at room temperature and as a function of their defect concentration are illustrated. The results are illustrated for the both symmetric and non-symmetric MoS$_2$ grain boundaries, as they were shown in Fig. 1. By increasing the defect concentration along the grain boundary, the thermal conductance is expected to decrease due to the increasing of the phonon-defect scattering rates. Nevertheless, based on our NEMD results, this observation appears to be valid only for the symmetric grain boundaries. In this regard, for the all considered MoS$_2$ grain boundaries consisting of 5-7, 6-8 and 4-6 dislocation cores, the thermal conductance of non-symmetric grain boundaries are found to be lower than the symmetric grain boundaries with higher defect concentration. This reveals higher phonon-phonon scattering rates along the non-symmetric grain boundaries, which originate from the mismatch of the phonon spectrum of the atoms in the two sides of the grain boundary. Nonetheless, and as expected, the symmetric grain boundaries with the lowest defect concentration present the highest thermal conductance. Based on our results, the thermal conductance of $MoS_2$ grain boundaries formed from 5-7 or 6-8 dislocation cores are very close. We found that $MoS_2$ armchair grain boundary with 4-8 dislocation core presents the minimum thermal conductance whereas the zigzag grain boundary formed from 4-8 defect pair yields distinctly higher thermal contact conductance (as shown in Fig. 2b). Interestingly, it was observed that the grain boundaries consisting of 4-6 dislocation core yield the strongest thermal conductance.

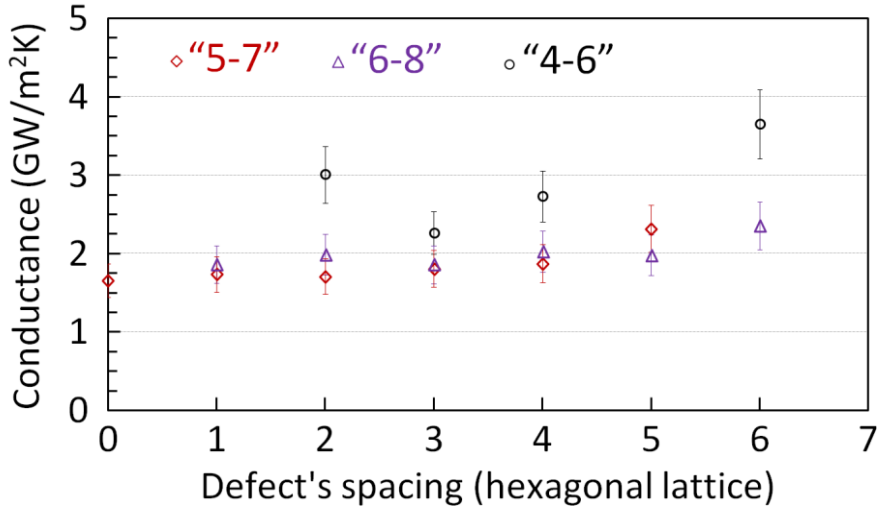

Fig. 3, $MoS_2$ grain boundaries thermal contact conductance predicted by the NEMD simulations. Here the results are shown for $MoS_2$ grain boundaries constructed from 5-7, 6-8 and 4-6 dislocation cores with different number of hexagonal lattices for defect's spacing. By increasing the defect's spacing, the defect concentration along the grain boundary decreases.

As discussed earlier, the acquired thermal contact conductance values were then used in our continuum modelling for the evaluation of effective thermal conductivity of macroscopic polycrystalline samples. In these modelling, for all grains, we considered an equal thermal conductivity of 37 W/mK, i.e. the one of pristine and single-crystalline $MoS_2$. For introducing the thermal conductance of every grain boundary in our FE modelling, we randomly selected a grain boundary and its corresponding thermal conductance was used.

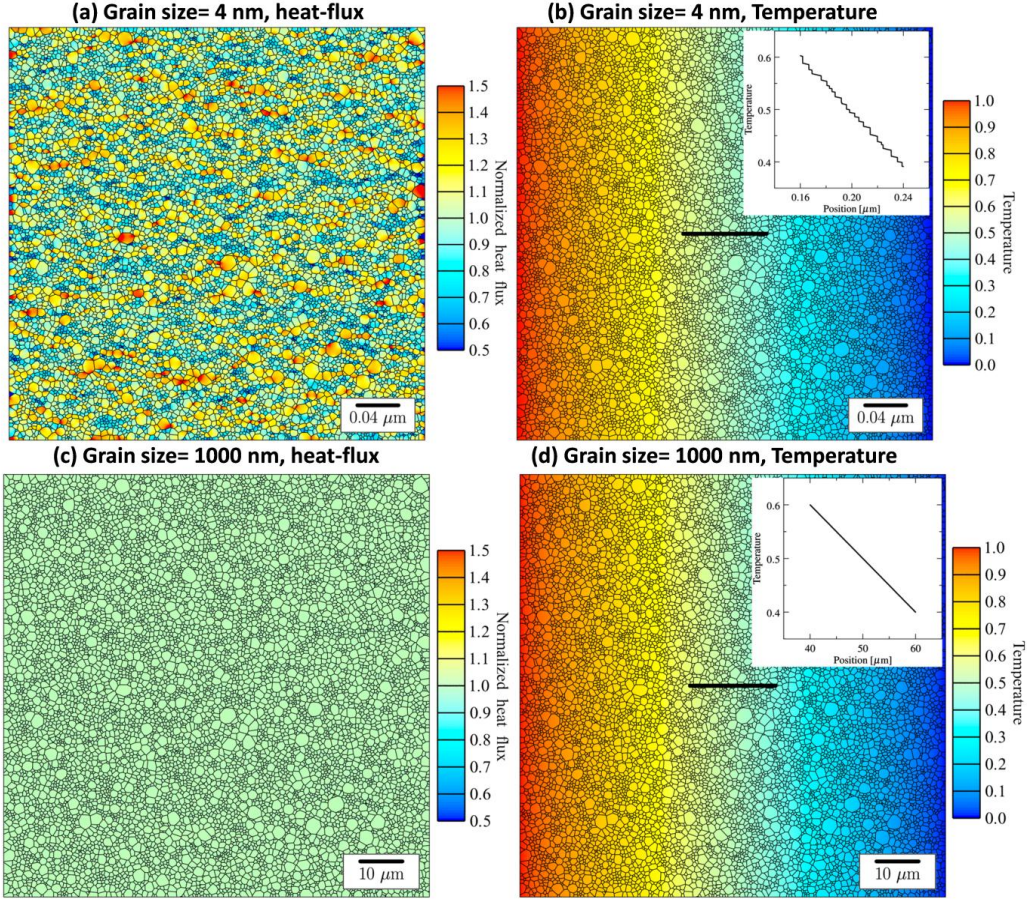

Fig. 4, NEMD/FE multiscale modeling results for the comparison of grain size effect on the established heat-flux and temperature distributions for polycrystalline MoS$_2$ with grain sizes of 4 nm and 1000 nm.

In Fig. 4, the NEMD/FE multiscale results for the calculated steady-state heat-flux and temperature profiles established along two single-layer polycrystalline MoS$_2$ with grain sizes of 4 nm and 1000 nm are compared. For the sample with equivalent grain size of 4 nm, it is clear that the grains are not equally involved in the heat-flux transfer (Fig. 4a). We note that the higher values for heat-flux illustrate the higher intensity of the thermal transport. In this case, because the conductance of the grain boundaries dominate the thermal transport, the heat-flux is transferred along the percolation paths in which there exist minimum resistance front the heat flow. Such that, the majority of the heat-flux are carried along grains that are large and elongated along the loading direction, percolating each other and preferably passing through grain boundaries with stronger thermal contact conductance. In addition, for polycrystalline films with small grain sizes, we found that the temperature distribution inside individual grains are almost constant, such that the temperature changes mainly occur at grain boundaries (Fig. 4b inset). On the contrary, for polycrystalline $MoS_2$ sheet with the large grain size of 1000 nm, the established temperature profile is uniform and the temperature gradient also form inside the grains (Fig. 4d inset), which implies that the effect of grain boundaries resistance is substantially weakened. In this case, different grains, irrespective of their sizes and their grain boundary configurations, are basically equally involved in the heat-flux transfer, which further confirms that the effect of grain boundaries resistance on the thermal transport is considerably weakened.

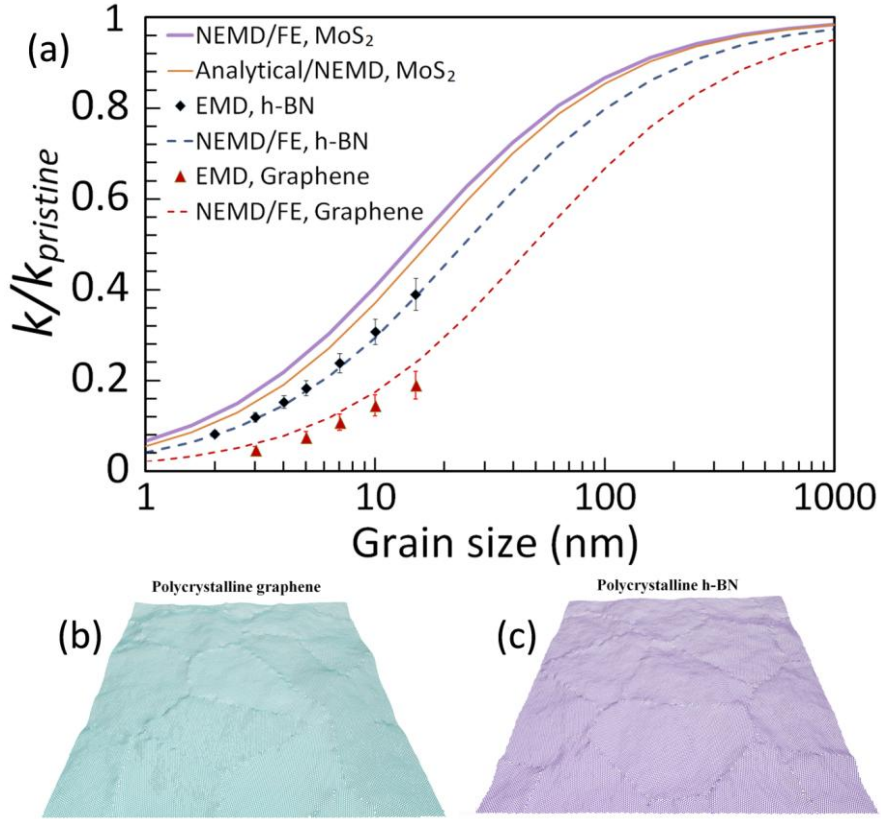

Fig. 5, NEMD/FE multiscale results for the normalized effective thermal conductivity of CVD grown polycrystalline $MoS_2$ as a function of grain size. The proposed NEMD/FE approach was also used to simulate the effective thermal conductivity of polycrystalline graphene and h-BN and the results for the normalized conductivity are compared with the results of fully atomistic models obtained using the equilibrium molecular dynamics (EMD) method. Atomistic models of polycrystalline (b) graphene and (c) h-BN with grain size of 15 nm. The EMD results for polycrystalline h-BN are based on our recent study [76] and similar methodology was used for polycrystalline graphene.

In Fig. 5, the normalized effective thermal conductivity of polycrystalline $MoS_2$ with respect to that of the pristine $MoS_2$ as a function of grain size, predicted by the NEMD/FE multiscale approach is illustrated. As it is shown, for polycrystalline samples with the grain sizes in nanoscale, the thermal conductivity initially increases sharply by increasing the grain size. By further increasing of the grain size and approaching the micro-scale, the increase in the thermal conductivity slow down and finally reaches a plateau, very close to the thermal conductivity of defect-free $MoS_2$. Since we assumed equal thermal conductivity for grains, the main factor that determines the thermal conductivity is the grain boundaries thermal conductance. Our modelling results reveal that the thermal conductivity of CVD grown $MoS_2$ films can be tuned by one order of magnitude through controlling the grain size. As discussed in our earlier study [53], the effective thermal conductivity of polycrystalline films can be also approximated by considering a series of line conductors connected by point thermal resistors. If one assumes constant values for the grains thermal conductivity ($k_{grain}$) and the grain boundary thermal conductance ($k_{GB}$), then the effective conductivity, $k_{eff}$, as a function of grain size, $L_{GB}$, can be obtained using the following relation [53]:

$$k_{eff} = \frac{k_{grain} \times k_{GB} \times L_{GB}}{k_{grain} + k_{GB} \times L_{GB}} \qquad (6)$$

To define the constant value for the grain boundary conductance in Eq. 6, we simply averaged the NEMD results for the constructed 20 $MoS_2$ grain boundaries. The results based on the mentioned analytical function are also compared with those that were acquired using our original FE polycrystalline modeling, as illustrated in Fig. 5. It can be seen that the analytical modeling distinctly underestimates the results of our elaborated FE modeling, especially for the polycrystalline samples with grain sizes smaller than 100 nm. As already discussed, for the ultra-fine grained $MoS_2$, the grain boundaries thermal conductance dominate the heat-flux transfer and such that considerable portions of the heat-flux are carried through percolation paths with minimal thermal resistances, which can consequently enhance the thermal transport. Such an enhancing effect on the thermal transport however cannot be considered in the 1D modelling and this well explain the underestimation of the FE results by the analytical function. As expected, by increasing the polycrystalline $MoS_2$ grain size, and subsequent decline of the percolation paths for the heat transfer, the results by the analytical function converge to those of FE modelling. Nevertheless, our study highlights that the simple analytical function based on the line and point conductors can be considered as a lower-bound to estimate the effective thermal conductivity along the polycrystalline materials.

In our earlier work [76], we used equilibrium molecular dynamics (EMD) for the direct evaluation of the thermal conductivity of polycrystalline h-BN using the fully atomistic models. However, construction of fully atomistic models of polycrystalline $MoS_2$ is more complicated, because their grain boundary configurations are much more complex than those for graphene or h-BN. To further evaluate the accuracy of the proposed NEMD/FE modeling in this study, we compared our results with those based on the EMD method for fully atomistic models for graphene and h-BN polycrystals. To do so, we extended our recent work for the polycrystalline h-BN [76] to the polycrystalline graphene. The optimized Tersoff potentials proposed by Lindsay and Broido were used for the modelling of graphene [77] and h-BN [78]. We also additionally constructed polycrystalline graphene and h-BN with the grain size of 15 nm with around 88,000 atoms, as shown in Fig. 5b and Fig. 5c, respectively. In our most recent study [79], we also used the NEMD method to estimate the thermal contact conductance of 6 different grain boundaries for the both graphene and h-BN. This way, we also used the NEMD results for the graphene and h-BN grain boundaries in our constructed FE model with 10,000 grains. In these cases, the thermal conductivity of graphene and h-BN grains were considered to be 3000 W/mK and 600 W/mK, respectively, according to our NEMD results for their single-crystalline films. In Fig. 5, we compared the normalized thermal conductivity of polycrystalline graphene and h-BN predicted by the proposed NEMD/FE approach and calculated using the EMD method for fully atomistic structures. Interestingly, the predictions by the NEMD/FE and EMD methods match accurately for the normalized thermal conductivity of polycrystalline h-BN. Nevertheless, the NEMD/FE slightly overestimates the EMD results for the polycrystalline graphene. It should be however noted that as discussed in a recent study [80], the EMD method implemented in the LAMMPS underestimates the thermal conductivity of graphene. Our EMD calculations predict the thermal conductivity of pristine graphene and h-BN around 1000±100 W/mK and 300±20 W/mK, respectively, which are considerably smaller than our NEMD results. However, comparison with the NEMD/FE results clearly reveals that the EMD method can acceptably simulate the relative reduction in the thermal conductivity of polycrystalline films, though it underestimates the values for the thermal conductivity. Nonetheless, the NEMD/FE method can be considered as a reliable approach not only to study the relative reduction but also more importantly to report exact values for the thermal conductivity (provided that the thermal conductivity of single-crystalline film is reproduced accurately). Our multiscale modelling results reveal (shown in Fig. 5) that the thermal conductivity of polycrystalline $MoS_2$, graphene or h-BN can be tuned by one order of magnitude through controlling the grain size. However, based on our findings this goal can be achieved only by fabrication of polycrystalline nanomembranes with grain sizes smaller than 100 nm. For the 2D materials with ultra-fine grain sizes, we should nevertheless remind that the electronic properties may substantially get affected [50,56,58], which may not be desirable for many applications. Worthy to mention that in the case of graphene nanomembranes, experimental studies have shown the possibility of tuning the thermal conductivity by defect engineering [81,82] or fabrication of graphene laminates [83,84]. These experimental routes [81–84] can be also extended for the tuning of thermal transport along TMDs nanomembranes.

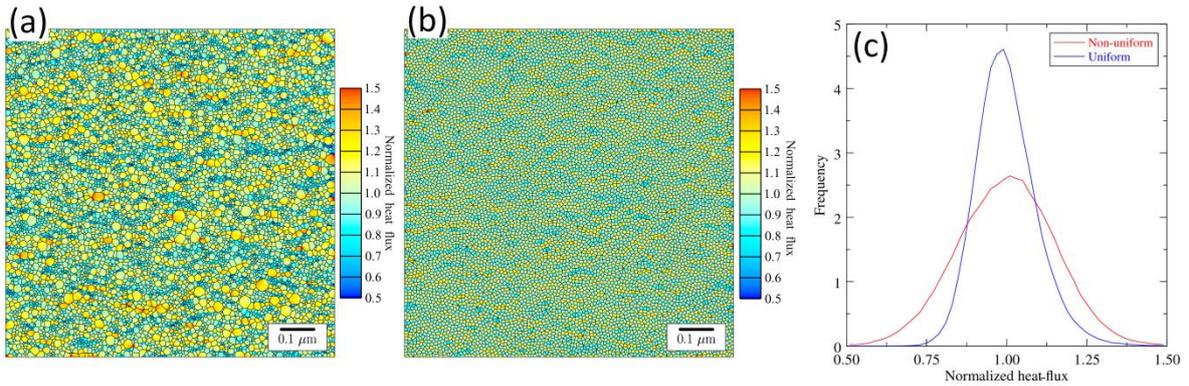

Fig. 6, NEMD/FE multiscale modeling results for the comparison of grain size distribution effect on the established heat-flux for polycrystalline $MoS_2$ with the grain size of 10 nm. Heat-flux distribution for (a) non-uniform and (b) uniform grain size configurations. (c) Heat-flux distribution function of uniform (blue line) compared to the non-uniform (red line) grains configurations.

To evaluate the roles of the grain size distribution on the established heat-flux field, a new microstructure was generated for which all grains have close grain sizes. This was achieved by using the same method as before, but in this case we imposed the close size to all grains (Dirac delta distribution). Two different grain size distributions were considered, one with a standard deviation of 0.45 (as before, see Section 2) and the other with a standard deviation of 0.01 (representing the uniform distribution). Since the effect of grain boundaries thermal conductance was found to be considerable for the grain sizes in nanoscale, in this case we accordingly studied the polycrystalline $MoS_2$ with the small grain size of 10 nm. Fig. 6 provides the heat-flux distributions for these 2 microstructures with uniform and non-uniform grain size configurations. Comparing Fig. 6a and Fig. 6b, it appears that by reducing the grain size distribution (approaching to the uniform grain configuration), the heat-flux distribution becomes more homogenous and uniform. As discussed before, this was expected since largest grains provide stronger diffusion paths, especially when these grains form clusters along the diffusion direction. However, for the sample with uniform grain size (and therefore in the absence of preferential diffusion paths), heat-flux heterogeneity still is present, which is due to the distribution of different grain boundaries thermal conductance. Here, the relative influence of the grain size distribution and grain boundaries thermal conductance distribution can be quantified. Fig. 6c illustrates the heat-flux frequency distributions for the 2 considered microstructures, which both exhibit bell curve shapes but with different standard deviations. As the grain size distribution changes from non-uniform to uniform, the standard deviation is reduced from 0.156 to 0.096. This simply means that the grain size distribution is responsible for 1/3 of the heat-flux variations while the grain boundaries thermal conductance distribution is responsible for 2/3 of the heat-flux variations. For the $MoS_2$ polycrystalline samples with grain sizes in micrometer, the effect of grain boundaries thermal conductance get substantially suppressed and therefore the effective thermal conductivity insignificantly depends on the grain size configurations [76].

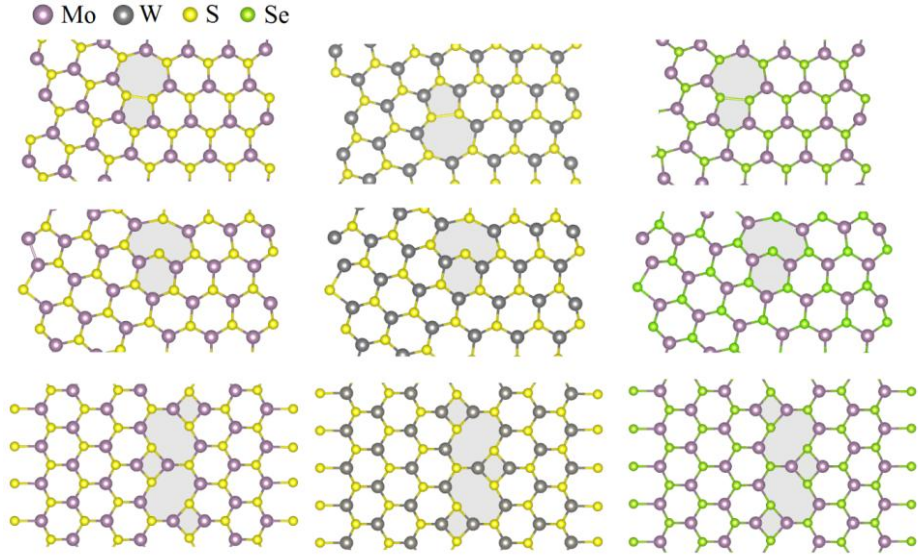

Fig. 7, First principles density functional theory relaxation results for 5-7, 6-8 and 4-8 dislocation cores along various $MoS_2$, $WS_2$ and $MoSe_2$ grain boundaries.

We also used density functional theory (DFT) calculations using the Vienna ab initio simulation package (VASP) [85,86] along with the Perdew-Burke-Ernzerhof (PBE) generalized gradient approximation exchange-correlation functional [87]. In addition to the $MoS_2$, we constructed similar grain boundaries consisting of 5-7, 4-4, 4-6, 4-8 and 6-8, dislocation cores for $WS_2$, $MoSe_2$, $WSe_2$, $MoTe_2$ and $WTe_2$ and then we used conjugate gradient energy minimization method with $10^{-5}$ eV criteria for the relaxation of the structures. We found that for the all considered TMDs films, the constructed grain boundaries are stable (as shown in Fig. 7 for few samples). Interestingly, our first-principles investigation reveal that the trends for the mechanical strength of different grain boundaries are similar for the all considered TMDs single-layer films, which will be discussed in our oncoming study. Based on the observation of considerable similarities for the mechanical strength of TMDs grain boundaries, irrespective of their exact values for tensile strength, we also expect that their thermal conduction properties may also present similar trends. Therefore, even though our study only focused on polycrystalline $MoS_2$, our findings are probably extendable for the normalized thermal transport along other polycrystalline TMDs nanomembranes. As compared in Fig. 5, the thermal conductivity of CVD grown $MoS_2$ is closer to the pristine and single-layer $MoS_2$ as compared with polycrystalline graphene or h-BN. This can suggest that for TMDs polycrystalline films, the thermal conduction properties may not get considerably weakened and so their thermal management would not be substantially affected in comparison with their single-crystalline films.

4. Conclusion

We developed a combined non-equilibrium molecular dynamics (NEMD)/finite element (FE) multiscale modelling to explore the thermal transport along polycrystalline and single-layer $MoS_2$. In the proposed approach, we first employed the NEMD method to investigate the thermal conductivity of pristine and defect-free $MoS_2$ at room temperature. In particular, extensive NEMD simulations with a reactive forcefield were carried out to evaluate interfacial thermal conductance of 20 different grain boundaries, that are observable in experimental samples of CVD grown $MoS_2$. In order to explore the effective thermal conductivity of samples at macroscopic level, we constructed continuum models of polycrystalline $MoS_2$ using the finite element approach. The acquired results based on the NEMD simulations were employed to introduce grains thermal conductivity as well as thermal conductance between every contacting grains. Our multiscale modelling results reveal that the thermal conductivity of CVD grown MoS$_2$ films can be tuned by one order of magnitude through controlling their grain size. Remarkably, by comparing the normalized thermal conductivity of CVD grown MoS$_2$, graphene and h-BN, it was found that the thermal transport along the MoS$_2$ is less suppressed by the grain boundaries resistance. If one takes into consideration the MoS$_2$ as the representative member of transition metal dichalcogenides (TMDs) nanosheets, our results suggest that thermal conduction properties of TMDs polycrystalline films may not substantially decreases in comparison with their single-crystalline films. In response to the remarkably fast growth and interests in the field of 2D materials, the insight provided by this investigation can be very useful to efficiently explore the effective thermal conductivity of various 2D polycrystalline and heterostructures.

## Acknowledgment

BM and TR greatly acknowledge the financial support by European Research Council for COMBAT project (Grant number 615132). BM also thank Dr. Obaidur Rahaman for providing the DFT models of the relaxed structures. ACTvD and AO acknowledge funding from NSF grant #1462980.

## References

(1)  Novoselov, K. S.; Geim, A. K.; Morozov, S. V; Jiang, D.; Zhang, Y.; Dubonos, S. V; Grigorieva, I. V; Firsov, A. A. *Science* **2004**, *306* (5696), 666–669.

(2)  Geim, A. K.; Novoselov, K. S. *Nat. Mater.* **2007**, *6* (3), 183–191.

(3)  Schedin, F.; Geim, a K.; Morozov, S. V; Hill, E. W.; Blake, P.; Katsnelson, M. I.; Novoselov, K. S.; F. Schedin A.K. Geim, S. V. M. E. W. H. P. B. M. I. K. & K. S. N. *Nat. Mater.* **2007**, *6* (9), 652–655.

(4)  Castro Neto, A. H. .; Peres, N. M. R. .; Novoselov, K. S. .; Geim, A. K. .; Guinea, F. *Rev. Mod. Phys.* **2009**, *81* (1), 109–162.

(5)  de Sousa, J. M.; Botari, T.; Perim, E.; Bizao, R. A.; Galvao, D. S. *RSC Adv.* **2016**, *6* (80), 76915–76921.

(6)  Kubota, Y.; Watanabe, K.; Tsuda, O.; Taniguchi, T. *Science* **2007**, *317* (5840), 932–934.

(7)  Song, L.; Ci, L.; Lu, H.; Sorokin, P. B.; Jin, C.; Ni, J.; Kvashnin, A. G.; Kvashnin, D. G.; Lou, J.; Yakobson, B. I.; Ajayan, P. M. *Nano Lett.* **2010**, *10* (8), 3209–3215.

(8)  Thomas, A.; Fischer, A.; Goettmann, F.; Antonietti, M.; Müller, J.-O.; Schlögl, R.; Carlsson, J. M. *J. Mater. Chem.* **2008**, *18* (41), 4893.

(9)  Algara-Siller, G.; Severin, N.; Chong, S. Y.; Björkman, T.; Palgrave, R. G.; Laybourn, A.; Antonietti, M.; Khimyak, Y. Z.; Krasheninnikov, A. V.; Rabe, J. P.; Kaiser, U.; Cooper, A. I.; Thomas, A.; Bojdys, M. J. *Angew. Chemie - Int. Ed.* **2014**, *53* (29), 7450–7455.

(10) Aufray, B.; Kara, A.; Vizzini, Ś .; Oughaddou, H.; Ĺ andri, C.; Ealet, B.; Le Lay, G. *Appl. Phys. Lett.* **2010**, *96* (18).

(11) Vogt, P.; De Padova, P.; Quaresima, C.; Avila, J.; Frantzeskakis, E.; Asensio, M. C.; Resta, A.; Ealet, B.; Le Lay, G. *Phys. Rev. Lett.* **2012**, *108* (15).

(12) Das, S.; Demarteau, M.; Roelofs, A. *ACS Nano* **2014**, *8* (11), 11730–11738.

(13) Li, L.; Yu, Y.; Ye, G. J.; Ge, Q.; Ou, X.; Wu, H.; Feng, D.; Chen, X. H.; Zhang, Y. *Nat. Nanotechnol.* **2014**, *9* (5), 372–377.

(14) Mannix, A. J.; Zhou, X.-F.; Kiraly, B.; Wood, J. D.; Alducin, D.; Myers, B. D.; Liu, X.; Fisher, B. L.; Santiago, U.; Guest, J. R.; Yacaman, M. J.; Ponce, A.; Oganov, A. R.; Hersam, M. C.; Guisinger, N. P. *Science (80-. ).* **2015**, *350* (6267), 1513–1516.

(15) Bianco, E.; Butler, S.; Jiang, S.; Restrepo, O. D.; Windl, W.; Goldberger, J. E. *ACS Nano* **2013**, *7* (5), 4414–4421.

(16) Zhu, F.; Chen, W.; Xu, Y.; Gao, C.; Guan, D.; Liu, C. *arXiv* **2015**, 1–20.

(17) Wang, Y.; Ding, Y. *Solid State Commun.* **2013**, *155*, 6–11.

(18) Radisavljevic, B.; Radenovic, A.; Brivio, J.; Giacometti, V.; Kis, A. *Nat. Nanotechnol.* **2011**, *6* (3), 147–150.

(19) Lee, C.; Wei, X.; Kysar, J. W.; Hone, J. *Science (80-. ).* **2008**, *321* (18 July 2008), 385–388.

(20) Balandin, A. A. *Nat. Mater.* **2011**, *10* (8), 569–581.

(21) Wang, Q. H.; Kalantar-Zadeh, K.; Kis, A.; Coleman, J. N.; Strano, M. S. *Nat. Nanotechnol.* **2012**, *7* (11), 699–712.

(22) He, Z.; Que, W. *Applied Materials Today.* 2016, pp 23–56.

(23) Reale, F.; Sharda, K.; Mattevi, C. *Appl. Mater. Today* **2016**, *3*, 11–22.

(24) Li, S.; Wang, S.; Tang, D. M.; Zhao, W.; Xu, H.; Chu, L.; Bando, Y.; Golberg, D.; Eda, G. *Appl. Mater. Today* **2015**, *1* (1), 60–66.

(25) Ambrosi, A.; Sofer, Z.; Pumera, M. *Small* **2015**, *11* (5), 605–612.

(26) Ambrosi, A.; Pumera, M. *ACS Catal.* **2016**, *6* (6), 3985–3993.

(27) Presolski, S.; Pumera, M. *Mater. Today* **2016**, *19* (3), 140–145.

(28) Eftekhari, A. *Appl. Mater. Today* **2017**, *8*, 1–17.

(29) Liu, Y.; Peng, X. *Appl. Mater. Today* **2017**, *7*, 1–12.

(30) Li, X.; Cai, W.; An, J.; Kim, S.; Nah, J.; Yang, D.; Piner, R.; Velamakanni, A.; Jung, I.; Tutuc, E.; Banerjee, S. K.; Colombo, L.; Ruoff, R. S. *Science (80-. ).* **2009**, *324* (5932), 1312–1314.

(31) Lin, Y.-M.; Valdes-Garcia, A.; Han, S.-J.; Farmer, D. B.; Meric, I.; Sun, Y.; Wu, Y.; Dimitrakopoulos, C.; Grill, A.; Avouris, P.; Jenkins, K. A. *Science (80-. ).* **2011**, *332* (6035), 1294–1297.

(32) Huang, P. Y.; Ruiz-Vargas, C. S.; van der Zande, A. M.; Whitney, W. S.; Levendorf, M. P.; Kevek, J. W.; Garg, S.; Alden, J. S.; Hustedt, C. J.; Zhu, Y.; Park, J.; McEuen, P. L.; Muller, D. a. *Nature* **2011**, *469* (7330), 389–392.

(33) Kim, K. K.; Hsu, A.; Jia, X.; Kim, S. M.; Shi, Y.; Hofmann, M.; Nezich, D.; Rodriguez-Nieva, J. F.; Dresselhaus, M.; Palacios, T.; Kong, J. *Nano Lett.* **2012**, *12* (1), 161–166.

(34) Wen, Y.; Shang, X.; Dong, J.; Xu, K.; He, J.; Jiang, C. *Nanotechnology* **2015**, *26* (27), 275601.

(35) Lee, K. H.; Shin, H. J.; Lee, J.; Lee, I. Y.; Kim, G. H.; Choi, J. Y.; Kim, S. W. *Nano Lett.* **2012**, *12* (2), 714–718.

(36) Gibb, A. L.; Alem, N.; Chen, J. H.; Erickson, K. J.; Ciston, J.; Gautam, A.; Linck, M.; Zettl, A. *J. Am. Chem. Soc.* **2013**, *135* (18), 6758–6761.

(37) Ramakrishna Matte, H. S. S.; Gomathi, A.; Manna, A. K.; Late, D. J.; Datta, R.; Pati, S. K.; Rao, C. N. R. *Angew. Chemie - Int. Ed.* **2010**, *49* (24), 4059–4062.

(38) Zhang, W.; Huang, J. K.; Chen, C. H.; Chang, Y. H.; Cheng, Y. J.; Li, L. J. *Adv. Mater.* **2013**, *25* (25), 3456–3461.

(39) Cong, C.; Shang, J.; Wu, X.; Cao, B.; Peimyoo, N.; Qiu, C.; Sun, L.; Yu, T. *Adv. Opt. Mater.* **2014**, *2* (2), 131–136.

(40) Gao, Y.; Liu, Z.; Sun, D.-M.; Huang, L.; Ma, L.-P.; Yin, L.-C.; Ma, T.; Zhang, Z.; Ma, X.-L.; Peng, L.-M.; Cheng, H.-M.; Ren, W. *Nat. Commun.* **2015**, *6*, 8569.

(41) Huang, C.; Wu, S.; Sanchez, A. M.; Peters, J. J. P.; Beanland, R.; Ross, J. S.; Rivera, P.; Yao, W.; Cobden, D. H.; Xu, X. *Nat. Mater.* **2014**, *13* (12), 1–6.

(42) Eichfeld, S. M.; Hossain, L.; Lin, Y.; Piasecki, A. F.; Kupp, B.; Birdwell, A. G.; Burke, R. A.; Lu, N.; Peng, X.; Li, J.; Azcatl, A.; McDonnell, S.; Wallace, R. M.; Kim, M. J.; Mayer, T. S.; Redwing, J. M.; Robinson, J. A. *ACS Nano* **2015**, *9* (2), 2080–2087.

(43) van der Zande, A. M.; Huang, P. Y.; Chenet, D. a; Berkelbach, T. C.; You, Y.; Lee, G.-H.; Heinz, T. F.; Reichman, D. R.; Muller, D. a; Hone, J. C. *Nat. Mater.* **2013**, *12* (6), 554–561.

(44) Najmaei, S.; Liu, Z.; Zhou, W.; Zou, X.; Shi, G.; Lei, S.; Yakobson, B. I.; Idrobo, J.-C.; Ajayan, P. M.; Lou, J. *Nat. Mater.* **2013**, *12* (8), 754–759.

(45) Zhou, W.; Zou, X.; Najmaei, S.; Liu, Z.; Shi, Y.; Kong, J.; Lou, J.; Ajayan, P. M.; Yakobson, B. I.; Idrobo, J. C. *Nano Lett.* **2013**, *13* (6), 2615–2622.

(46) Gong, Y.; Lin, J.; Wang, X.; Shi, G.; Lei, S.; Lin, Z.; Zou, X.; Ye, G.; Vajtai, R.; Yakobson, B. I.; Terrones, H.; Terrones, M.; Tay, B. K.; Lou, J.; Pantelides, S. T.; Liu, Z.; Zhou, W.; Ajayan, P. M. *Nat Mater* **2014**, *13* (12), 1135–1142.

(47) Tongay, S.; Fan, W.; Kang, J.; Park, J.; Koldemir, U.; Suh, J.; Narang, D. S.; Liu, K.; Ji, J.; Li, J.; Sinclair, R.; Wu, J. *Nano Lett.* **2014**, *14* (6), 3185–3190.

(48) Yoo, Y.; Degregorio, Z. P.; Johns, J. E. *J. Am. Chem. Soc.* **2015**, *137* (45), 14281–14287.

(49) Chen, K.; Wan, X.; Wen, J.; Xie, W.; Kang, Z.; Zeng, X.; Chen, H.; Xu, J. Bin. *ACS Nano* **2015**, *9* (10), 9868–9876.

(50) Roche, A. I. and A. W. C. and L. C. and L. C. and J. M. K. and S. *2D Mater.* **2017**, *4* (1), 12002.

(51) Rasool, H. I.; Ophus, C.; Klug, W. S.; Zettl, a.; Gimzewski, J. K. *Nat. Commun.* **2013**, *4*, 2811.

(54) Aksamija, Z.; Knezevic, I. *Phys. Rev. B - Condens. Matter Mater. Phys.* **2014**, *90* (3).

(55) Hahn, K. R.; Melis, C.; Colombo, L. *Carbon N. Y.* **2016**, *96*, 429–438.

(56) Van Tuan, D.; Kotakoski, J.; Louvet, T.; Ortmann, F.; Meyer, J. C.; Roche, S. *Nano Lett.* **2013**, *13* (4), 1730–1735.

(57) Garrido, M. S. and J. E. B. V. and M. B. and M. S. and A. W. C. and S. R. and J. A. *2D Mater.* **2015**, *2* (2), 24008.

(58) Cummings, A. W.; Duong, D. L.; Nguyen, V. L.; Van Tuan, D.; Kotakoski, J.; Barrios Vargas, J. E.; Lee, Y. H.; Roche, S. *Adv. Mater.* **2014**, *26* (30), 5079–5094.

(59) Renteria, J.; Samnakay, R.; Rumyantsev, S. L.; Jiang, C.; Goli, P.; Shur, M. S.; Balandin, A. A. *Appl. Phys. Lett.* **2014**, *104* (15).

(60) Jiang, C.; Samnakay, R.; Rumyantsev, S. L.; Shur, M. S.; Balandin, A. A. *Appl. Phys. Lett.* **2015**, *106* (2).

(61) Jiang, C., S. L. Rumyantsev, R. Samnakay, M. S. Shur, A. A. B. *J. Appl. Phys.* **2015**, *117* (6), 64301.

(62) Plimpton, S. *J. Comput. Phys.* **1995**, *117* (1), 1–19.

(63) Ostadhossein, A.; Rahnamoun, A.; Wang, Y.; Zhao, P.; Zhang, S.; Crespi, V. H.; van Duin, A. C. T. *J. Phys. Chem. Lett.* **2017**, *8* (3), 631–640.

(67) Lautensack, C.; Zuyev, S. *Adv. Appl. Probab.* **2008**, *40* (3), 630–650.

(68) Torres, M. S. and B. G. and M. P. and D. S. R. and A. E. S. and J. S. R. and F. A. and B. M. and R. Q. and L. C. and S. R. and C. M. S. *2D Mater.* **2016**, *3* (3), 35016.

(69) Chaboche, J. L.; Cailletaud, G. *Comput. Methods Appl. Mech. Eng.* **1996**, *133* (1), 125–155.

(70) Cailletaud, G.; Forest, S.; Jeulin, D.; Feyel, F.; Galliet, I.; Mounoury, V.; Quilici, S. *Comput. Mater. Sci.* **2003**, *27* (3), 351–374.

(71) Geers, M. G. D.; Cottura, M.; Appolaire, B.; Busso, E. P.; Forest, S.; Villani, A. *J. Mech. Phys. Solids* **2014**, *70* (1), 136–153.

(72) Schelling, P. K.; Phillpot, S. R.; Keblinski, P. *Phys. Rev. B* **2002**, *65* (14), 1–12.

(73) Yan, R.; Simpson, J. R.; Bertolazzi, S.; Brivio, J.; Watson, M.; Wu, X.; Kis, A.; Luo, T.; Hight Walker, A. R.; Xing, H. G. *ACS Nano* **2014**, *8* (1), 986–993.

(74) Sahoo, S.; Gaur, A. P. S.; Ahmadi, M.; Guinel, M. J. F.; Katiyar, R. S. *J. Phys. Chem. C* **2013**, *117* (17), 9042–9047.

(75) Liu, J.; Choi, G. M.; Cahill, D. G. *J. Appl. Phys.* **2014**, *116* (23).

(78) Lindsay, L.; Broido, D. a. *Phys. Rev. B* **2011**, *84* (15), 1–6.

(80) Fan, Z.; Pereira, L. F. C.; Wang, H.-Q.; Zheng, J.-C.; Donadio, D.; Harju, A. *Phys. Rev. B* **2015**, *92* (9), 94301.

(81) Malekpour, H.; Ramnani, P.; Srinivasan, S.; Balasubramanian, G.; Nika, D. L.; Mulchandani, A.; Lake, R. K.; Balandin, A. A. *Nanoscale* **2016**, *8* (30), 14608–14616.

(82) Malekpour, H.; Ramnani, P.; Srinivasan, S.; Balasubramanian, G.; Nika, D. L.; Mulchandani, A.; Lake, R.; Balandin, A. A. **2016**, 23.

(83) Malekpour, H.; Chang, K.-H.; Chen, J.-C.; Lu, C.-Y.; Nika, D. L.; Novoselov, K. S.; Balandin, a a. *Nano Lett.* **2014**, *14* (9), 5155–5161.

(84) Renteria, J. D.; Ramirez, S.; Malekpour, H.; Alonso, B.; Centeno, A.; Zurutuza, A.; Cocemasov, A. I.; Nika, D. L.; Balandin, A. A. *Adv. Funct. Mater.* **2015**, *25* (29), 4664–4672.

(85) Kresse, G.; Furthm??ller, J. *Comput. Mater. Sci.* **1996**, *6* (1), 15–50.

(86) Kresse, G.; Furthmüller, J. *Phys. Rev. B* **1996**, *54* (16), 11169–11186.

(87) Perdew, J.; Burke, K.; Ernzerhof, M. *Phys. Rev. Lett.* **1996**, *77* (18), 3865–3868.